**Single-Crystal Growth and Physical Properties of the Antiferromagnetic Double Perovskite $Ba_2CoMoO_6$**

A.R.N. Hanna[1,*], M. M. Ferreira-Carvalho[2,3], S.H. Chen[2], C. F. Chang[2], C. Y. Kuo[4,5], A.T.M.N. Islam[1], R. Feyerherm[1], L.H. Tjeng[2], B. Lake[1,6,*]

[1]Helmholtz-Zentrum Berlin für Materialien und Energie GmbH, 14109 Berlin, Germany

[2]Max Planck Institute for Chemical Physics of Solids, Nöthnitzer Str. 40, 01187 Dresden, Germany

[3]Institute of Physics II, University of Cologne, Zülpicher Straße 77, 50937 Cologne, Germany

[4]Department of Electrophysics, National Yang Ming Chiao Tung University, 30010 Hsinchu, Taiwan

[5]National Synchrotron Radiation Research Center, 101 Hsin-Ann Road, 30076 Hsinchu, Taiwan

[6]Institut für Festkörperphysik, Technische Universität Berlin, Hardenbergstraße 36, 10623 Berlin, Germany

*Corresponding authors: Abanoub.Hanna@helmholtz-berlin.de, Bella.lake@helmholtz-berlin.de

## Abstract

This work presents a comprehensive investigation of the structural, magnetic, and electronic properties of the double perovskite $Ba_2CoMoO_6$ (BCMO). Single crystals were grown via floating-zone and Czochralski growth techniques and characterized using a set of complementary methods. X-ray diffraction analysis confirmed that BCMO crystallizes in a face-centered cubic (FCC) structure with space group *Fm-3m*. Magnetic susceptibility measurements exhibit antiferromagnetic ordering below $T_N = 20.1(1)$ K. A spin-flop transition is observed at 26.5 kOe. Heat capacity measurements and entropy analysis are consistent with a $J_{\mathrm{eff}} = ½$ ground state for $Co^{2+}$ ions. X-ray absorption spectroscopy provided insight into the local electronic structure, revealing the spin-orbit and crystal field effects. The cluster-model analysis yields a *g*-factor of 4.36, consistent with a spin-orbit-entangled ground state. Surface photovoltage spectroscopy demonstrates a strong near-surface optical response with an onset near 1.5 eV and a pronounced feature at 2.65 eV, consistent with visible-range Co–O–Mo transitions in line with related Ba-based double perovskites, highlighting its potential for spintronic and energy-conversion applications.

## Introduction

Double perovskite oxides with the general formula $A_2BB'O_6$ represent a versatile class of materials that exhibit diverse physical phenomena, including quantum magnetism, colossal magnetoresistance, and complex electronic behavior [1–5]. Their high-symmetry crystal structures—characterized by corner-sharing octahedra around alternating B and B′ cations—provide an ideal platform for tuning properties via chemical substitution at the A, B, or B′ sites [2,3,6–9]. When magnetic ions occupy the B site, they form a face-centred cubic (FCC) sublattice, where competing exchange interactions and geometric constraints can stabilize a wide range of ground states from simple ferromagnets to frustrated antiferromagnets with substantial short-range correlations. In many Co-based systems, high-spin $Co^{2+}$ in octahedral coordination realizes a $^4T_{1g}$ ground term with an effective orbital degree of freedom; spin–orbit coupling ($\lambda$) then splits this term into low-lying $J_{eff}$ = 1/2, 3/2, and 5/2 multiplets on the scale of $|\lambda|$, giving rise to spin-orbit-entangled $J_{eff}$ = 1/2 Kramers doublets that strongly influence magnetism and anisotropy [14–19,26].

Within this broader family, Ba-based double perovskites hosting Co together with *4d/5d* cations have attracted considerable interest as model systems where geometrical frustration, spin–orbit coupling, and electronic correlations can be disentangled [6–9,22,27]. $Ba_2CoMoO_6$ (BCMO; space group *Fm-3m*) is a particularly appealing member: $Co^{2+}$ and $Mo^{6+}$ form rock-salt-ordered $CoO_6$ and $MoO_6$ octahedra within a $Ba^{2+}$ framework, so that the Co ions occupy an undistorted FCC sublattice with well-defined Co–O–Mo–O–Co exchange pathways, closely analogous to the isovalent compound $Ba_2CoWO_6$ [10–12,27]. In contrast to $Sr_2FeMoO_6$, commonly described by $Fe^{3+}/Mo^{5+}$ valence states [13], stoichiometry and charge balance in BCMO point to a $Co^{2+}/Mo^{6+}$ valence combination, and the $Co^{2+}$ assignment can be tested directly by Co $L_{2,3}$-edge x-ray absorption spectroscopy (XAS) together with full-multiplet cluster calculations [14–21]. At the same time, antisite disorder between B and B' is known to strongly affect the magnetic and transport properties of double perovskites, as it suppresses the superlattice reflections associated with B-site order and disrupts the Co–O–Mo network; quantifying and minimizing such disorder is therefore essential for accessing the intrinsic properties, preferably in a single crystalline form.

However, conventional solid-state synthesis of $Ba_2CoMoO_6$ typically yields polycrystalline samples containing impurity phases such as $BaMoO_4$ and CoO, with the former arising particularly under air annealing due to $MoO_3$ volatility. Previous work on BCMO has been restricted to powders and has not resolved the intrinsic magnetic anisotropy, the detailed $J_{eff} = ½$ level scheme, or the near-surface optoelectronic response. Single crystals are required to overcome grain-boundary and texture effects, to quantify antisite disorder from superlattice reflections, and to establish a reliable link between local electronic structure and bulk magnetism. In this work, we grow high-purity BCMO single crystals via floating-zone (FZ) and Czochralski (CZ) methods under argon atmosphere, and combine X-ray diffraction, magnetometry, heat capacity, surface photovoltage spectroscopy, Co $L_{2,3}$-edge XAS, and X-ray photoelectron spectroscopy (XPS) for Mo 3d to build a coherent picture of its structural, magnetic, and electronic properties. We demonstrate cubic *Fm-3m* symmetry with rock-salt B-site order and no measurable Co/Mo antisite disorder within the sensitivity of powder diffraction, confirm a $Co^{2+}$ $J_{eff} = ½$ Kramers doublet ground state from thermodynamics and XAS, and show that the observed antiferromagnetic order and spin-flop transition are consistent with dominant second neighbour antiferromagnetic Co–O–Mo–O–Co superexchange interactions on the FCC lattice. Finally, we use surface photovoltage measurements to reveal a strong near-surface optical response with a prominent feature at 2.65 eV, placing BCMO in the broader landscape of Ba-based Mo/W double perovskites with visible-range optical transitions and suggesting potential spintronic and catalytic functionalities.

## Experimental Methods

### Powder Preparation

Polycrystalline BCMO was synthesized by two routes at the CoreLab Quantum Materials (HZB). First, a conventional solid-state reaction using stoichiometric $BaCO_3$, CoO, and $MoO_3$ was performed: the mixed powders were heated at 5 °C/min to 900 °C, calcined for 12 h, reground, and then heated at 5 °C/min to 1200 °C and sintered for 24 h, yielding BCMO with ≈5% secondary phases (mainly $BaMoO_4$ and CoO, as identified by powder x-ray diffraction (PXRD)). To improve phase purity, a sol-gel route was then adopted. Stoichiometric amounts (Ba:Co:Mo = 2:1:1) of $BaCO_3$ (99.99%), $Co(CH_3COO)_2$ (99.99%), and $(NH_4)_2MoO_4$ (prepared in-situ by reacting $MoO_3$ with aqueous $NH_3$) were dissolved in deionized water with oxalic acid as a chelating agent. The solution was heated on a hot plate in air at 250 °C to remove water and form a metal–oxalate precursor, which was then calcined in air at 2 °C/min to 500 °C for 6 h and to 800 °C for 10 h with an intermediate grinding. The resulting powder was pressed into pellets and annealed at 3 °C/min to 1200 °C for 12 h in flowing argon, producing black BCMO powder with a reduced impurity fraction of 3.74% (quantified by Rietveld refinement of PXRD). Owing to the volatility of $MoO_3$, phase purity was found to be highly sensitive to precursor stoichiometry, annealing temperature, and heating rate; therefore, heating profiles were carefully controlled and crucibles sealed to minimize contamination and Mo loss. The sol-gel argon-annealed powder was used as a precursor for crystal growth. Parts of the polycrystalline boules from CZ crystal growth were selected and used for XAS, X-ray photoelectron spectroscopy (XPS), surface photovoltage (SPV) and thermodynamic measurements.

## Crystal Growth

Single-crystal growth was first initiated using an optical floating-zone (FZ) furnace (Crystal Systems Corp., FZ-T-10000-H-VI-VPO) with four 300 W tungsten–halide lamps. Initial growths in air at ambient pressure showed that BCMO melts incongruently, decomposing into multiple immiscible phases (e.g., $Ba_2MoO_5$, $CoMoO_4$, as identified by PXRD), and did not yield stable single crystals. Subsequent trials under 2 bar Argon overpressure suppressed decomposition but still produced predominantly polycrystalline boules. Systematic improvements in the purity of the feed rods—via 2–3 cycles of annealing at 5 °C/min to 1200–1250 °C under argon for 24 h per cycle, each followed by slow cooling at 5 °C/min —reduced impurity levels below ≈ 3.74% and enhanced floating-zone stability, so that small single crystals (≈ 4×3×1 $mm^3$) could be cut from the as-grown rods. These results show that both argon overpressure and high-purity starting material are crucial to mitigate incongruent melting and preserve BCMO stoichiometry.

To further stabilize single-crystal growth, additional experiments were carried out using a Czochralski (CZ) furnace (Mini Czochralski Oxypuller 05-03, Cyberstar) equipped with radio frequency (RF) heating and an in-situ weighing system on the pulling rod. Growth was attempted under flowing argon at 0.5 bar overpressure with nominal pulling rates of 0.5–1.0 mm/h and crystal and melt rotation rates of 10–20 rpm. Under these conditions, the charge exhibited partially molten regions and persistent solid chunks, preventing the establishment of a stable melt zone for seeded growth. Nevertheless, slow cooling from the melt reproducibly yielded small single crystals (≈ 3×2.7×0.2 $mm^3$) with low impurity content. Laue backscattering patterns from these crystals exhibited sharp, well-defined spots consistent with the cubic symmetry of BCMO, confirming their single-crystalline nature.

## Characterization

The structural parameters and phase purity of the polycrystalline and single-crystalline BCMO samples were examined at room temperature by powder X-ray diffraction (PXRD) using a Bruker D8 diffractometer with Cu Kα radiation. The PXRD data were analyzed by Rietveld refinement using the *FullProf* suite to extract lattice parameters, atomic positions, and phase fractions, by refining the main phase in the cubic *Fm-3m* space group (No. 225) [23]. Magnetic properties were measured using a SQUID magnetometer in the temperature range 2–300 K and in applied magnetic fields up to 70 kOe (1 kOe = 0.1 T). Heat capacity was measured in a Physical Property Measurement System (PPMS, Quantum Design) between 2 and 300 K. Soft X-ray absorption spectroscopy (XAS) at the Co $L_{2,3}$-edges was carried out at beamline TPS45 the National Synchrotron Radiation Research Center (NSRRC), Hsinchu, Taiwan [34]. X-ray photoelectron spectroscopy (XPS) analyses were performed using a monochromatic Al Kα X-ray source (Focus 500, Specs), with a photon energy of 1400.00 eV, and an electron analyzer (Phoibos 100, Specs) at the institute of solar fuels, HZB (see SI).

The optical response was probed by surface photovoltage (SPV) spectroscopy at room temperature using the fixed-capacitor method with a mechanical chopper at 8 Hz for both single crystals and powders, over a photon energy range of 0.5–5 eV.

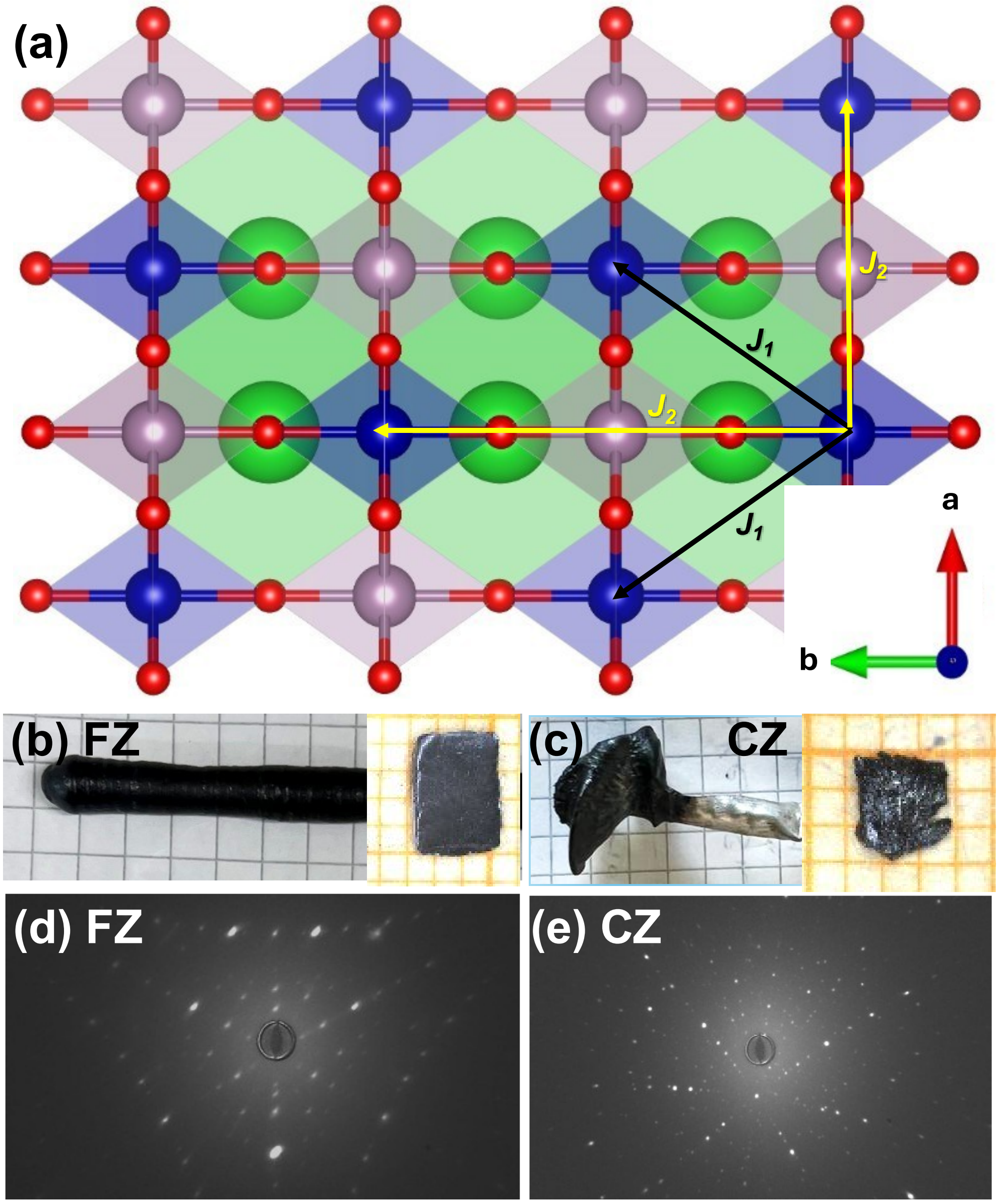

Figure 1. (a) The crystal structure of $Ba_2CoMoO_6$ (BCMO) consists of $Co^{2+}$ (blue), $O^{2-}$ (red), $Ba^{2+}$ (green) and $Mo^{6+}$ (Magenta) ions. The arrows represent the possible magnetic exchange interactions between $Co^{2+}$ ions: nearest neighbor interaction ($J_1$, black) and next nearest neighbor interaction ($J_2$, yellow) through superexchange between Co-O-Mo-O-Co. (b,c) The BCMO crystals, grown using: (b) the FZ technique, and (c) the CZ technique. The insets to the right show the separated single crystals of dimensions 4×3×1 $mm^3$ by FZ and 3×2.7×0.2 $mm^3$ by CZ. (d, e) Laue diffraction patterns of these crystals from FZ and CZ growths, showing [111] and [100] symmetry, respectively.

**Results**

**Structural characterization**

PXRD analysis confirmed that BCMO crystallizes in an FCC structure with space group *Fm-3m* (225) and regular undistorted rock-salt order consisting of regular $BaO_{12}$ cuboctahedra and regular corner-sharing $CoO_6$ and $MoO_6$ octahedra. $Ba^{2+}$ occupies the *8c* Wyckoff position, while $Co^{2+}$ occupies the *4b* position, $Mo^{6+}$ occupies the *4a* position and $O^{2-}$ occupies the *24e* position. All these atomic positions are fixed by the space group symmetry except for oxygen.

Antisite disorder is a key parameter in double perovskites because it suppresses the intensity of the superlattice reflections associated with B-site order [13]. In BCMO, we therefore allowed the Ba, Co, and Mo occupancies to vary during refinement; they do not deviate significantly from ideal stoichiometry. The PXRD pattern, including the superlattice reflections, and the Rietveld refinements are consistent with the ordered *Fm–3m* structure, with $Co^{2+}$ and $Mo^{6+}$ occupying the expected crystallographic sites and no measurable antisite disorder within the sensitivity of the measurement (1-2%). The refined oxygen occupancy is consistent with a stoichiometric composition within the limits of PXRD. These results are also supported by XPS spectrum for BCMO which indicates the absence of $Mo^{5+}$ in the polycrystalline sample (Additional details on the analysis of magnetic susceptibility and XPS measurements are provided in the Supplemental Material [53]).

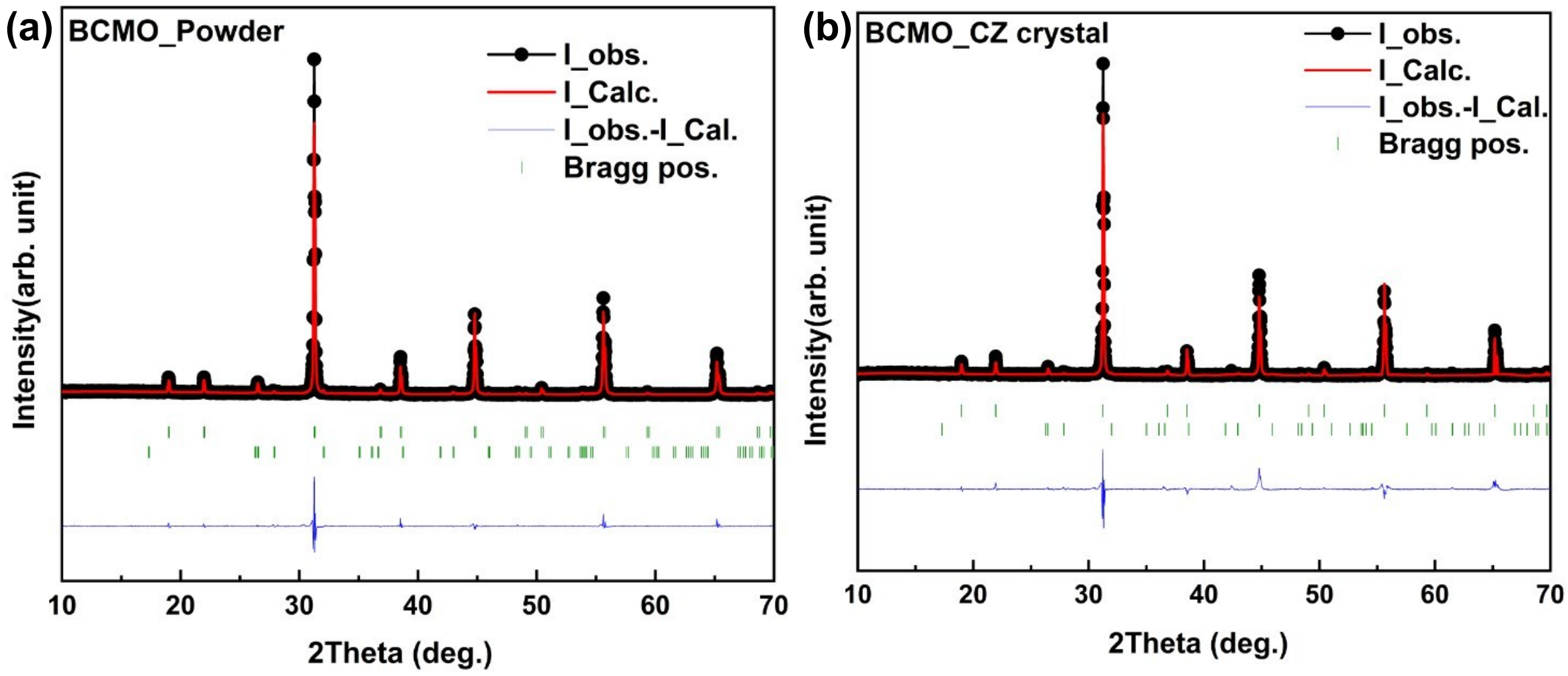

Figure 2. Rietveld refinement of the X-ray powder diffraction of BCMO: (a) Sol-gel annealed powder with 3.74% $BaMoO_4$, (b) CZ crystal with 2.31% $BaMoO_4$. The data is represented by the black circles, while the refinement is given by the red line which included both the BCMO (Green-Top ticks) and $BaMoO_4$(Green-bottom ticks) phases. The blue line gives the difference between fit and data.

Figure 2 shows the refinement of the PXRD patterns of different samples of BCMO. The refinement factors are summarized for the sol-gel Argon-annealed powder and both the CZ and FZ crystals of BCMO in Table 1. Both polycrystalline samples, prepared by solid-state synthesis and the sol-gel method, were annealed in argon and contained a significant amount of impurities (≈ 3.74 %). The structural refinement suggested that the main impurity phase is $BaMoO_4$ with space group *I 41/a (*88). Nevertheless, the argon-annealed powder has a slightly smaller amount of impurities than the reported powder (4.5%) [22], which was prepared in oxygen. Both crystals grown by the CZ and FZ methods have almost the same amount of impurity (2.33(2) %).

Table 1. Rietveld Refinement parameters for BCMO samples, including the lattice parameter (*a*), the *R* factors including the profile *R*-factor ($R_p$), weighted-profile *R*-factor ($R_{wp}$), and expected *R*-factor ($R_{exp}$), the atomic positions of oxygen in the *x* direction ($_x$(**O**)) and the occupancy of oxygen ($\mathbf{O}_{occ}$)

| | **Impurities (%) $BaMoO_4$** | ***a*** (Å) | $R_p$ (%) | $R_{wp}$ (%) | $R_{exp}$ (%) | ***Chi²*** | ***x(O)*** | $O_{occ}$ |
|---|---|---|---|---|---|---|---|---|
| **Sol-gel Annealed** | 3.74(3) | 8.088 | 3.99 | 5.93 | 2.65 | 5.01 | 0.26 | 1 |
| **Crystal-CZ** | 2.31(2) | 8.088 | 5.16 | 7.81 | 2.81 | 7.70 | 0.25 | 1 |
| **Crystal-FZ** | 2.35(2) | 8.088 | 4.25 | 8.60 | 3.75 | 8.00 | 0.25 | 1 |
| **Reported [22]** | 4.5 | 8.086 | 4.26 | 5.76 | 3.69 | 2.43 | 0.25 | 1 |

## Magnetic Properties

Figure 3 shows the temperature dependence of the magnetic susceptibility $\chi(T)$ and its inverse $1/\chi(T)$ for the polycrystalline sample and for single crystals oriented along [100] and [111], measured at 0.3, 1, and 10 kOe. The polycrystalline sample exhibits a pronounced concave curvature in $1/\chi(T)$ , probably due to a higher $BaMoO_4$ impurity fraction than the single crystals, as confirmed by powder X-ray diffraction. Due to the difficulty in obtaining large, high-quality single crystals of BCMO, orientation-resolved measurements were performed on separate crystals grown by different methods: a [111]-oriented crystal from FZ growth in argon and a [100]-oriented crystal from CZ growth in argon. These different growth conditions lead to small but systematic differences in crystal quality and magnetic response, with the [100] crystal exhibiting an additional high-temperature feature around 300 K at high fields, consistent with residual impurities below the PXRD detection limit.

For the remainder of the quantitative magnetic analysis, we therefore focus on the [111] FZ crystal, which shows the cleanest intrinsic behavior as revealed by sharp Laue patterns, the absence of a large diamagnetic offset, and the lowest impurity content. All samples, regardless of morphology or growth method, exhibit the same Néel temperature of $T_N$ = 20.1(1) K. The inverse susceptibility of all samples deviates strongly from linear Curie–Weiss behavior between about 30 and 150 K, closely resembling the response reported for BCWO and demonstrating that a simple Curie-Weiss law is not adequate [15,24,25]. This nonlinearity is naturally understood in terms of thermal population of low-lying excited states within the $Co^{2+}$ $^4T_1$ multiplet in an octahedral crystal field with substantial spin–orbit coupling, rather than being attributable solely to magnetic interactions or extrinsic defects [25]. For high spin octahedral $Co^{2+}$ ($3d^7$), the $^4T_{1g}$ ground term carries an effective orbital angular momentum $l$ = 1; spin–orbit coupling λ, splits this triplet into low-lying $J_{\mathrm{eff}}$ = 1/2, 3/2, and 5/2 multiplets separated by energies of order |λ|. This level scheme and the associated low lying excited states are well established for octahedral $Co^{2+}$ complexes [15,26].

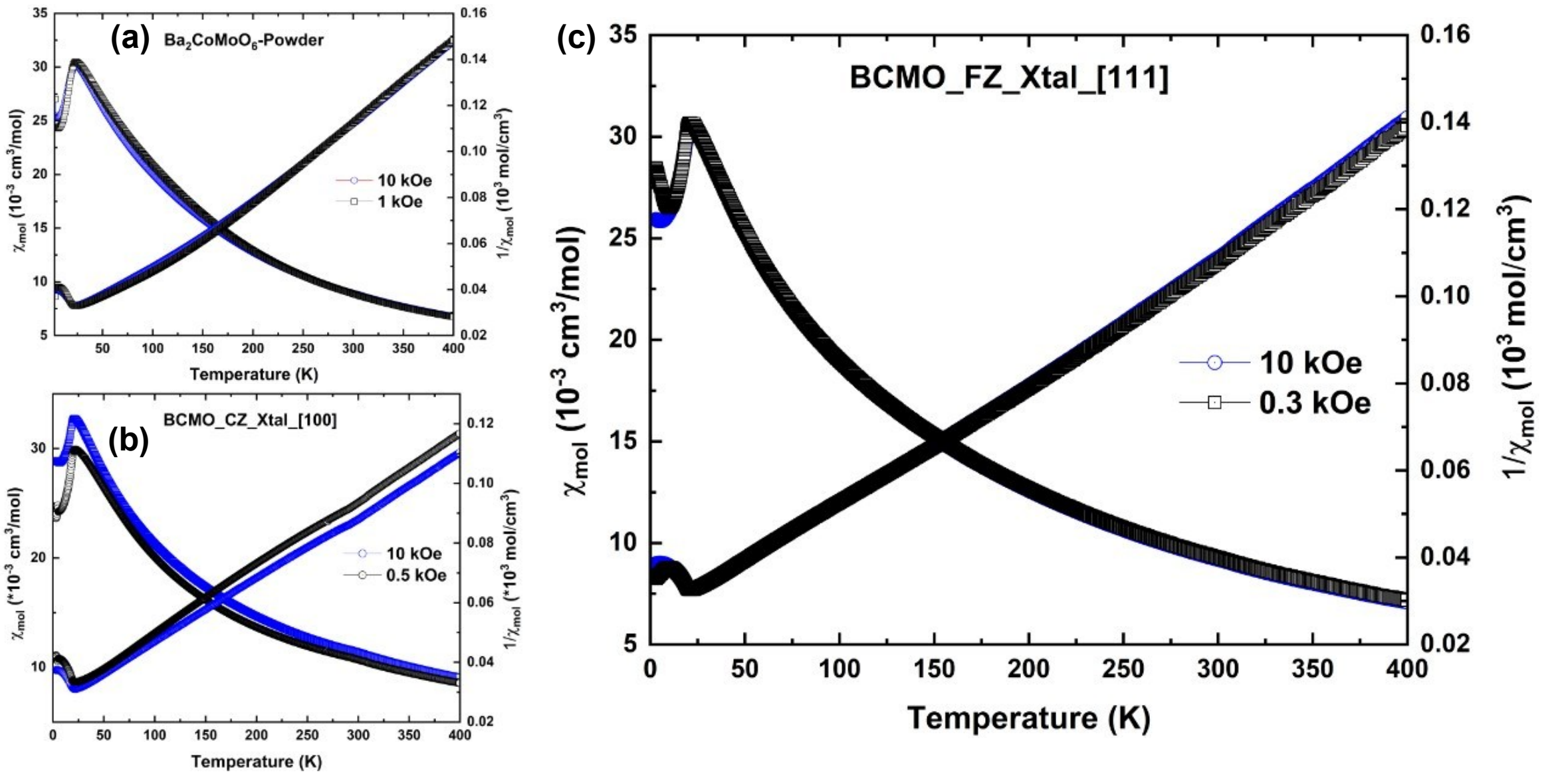

Figure 3. Temperature dependence of magnetic susceptibility (left $y$-axis) and inverse susceptibility (right $y$-axis) of BCMO samples at 0.3 and 10 kOe: (a) polycrystalline powder, (b) CZ crystal with field applied along the [100] direction, and (c) FZ crystal with field along the [111] direction.

To describe the $Co^{2+}$ magnetic response quantitatively, the susceptibility of BCMO is simulated for the fields applied using a cluster model with parameters derived from the XAS analysis discussed later. The antiferromagnetic interactions between the Co ions are considered, using Weiss' molecular field approach $\chi = M/(H + wM)$, The fitted molecular-field parameter $w$ accounts for the antiferromagnetic Co–Co interactions and produces a positive offset in the inverse susceptibility. For both applied fields we obtain a very good match between experiment and calculation for $T > 100$ K, see Fig. 4, with $w = 20$ mol/cm³ at 0.3 kOe and $w = 21$ mol/cm³ at 10 kOe. In the simple mean-field limit, $w$ can be expressed as an effective Weiss temperature through $\theta_W = -Cw$, giving $\theta_W \approx -37$ K which indicates a weak frustration in BCMO.

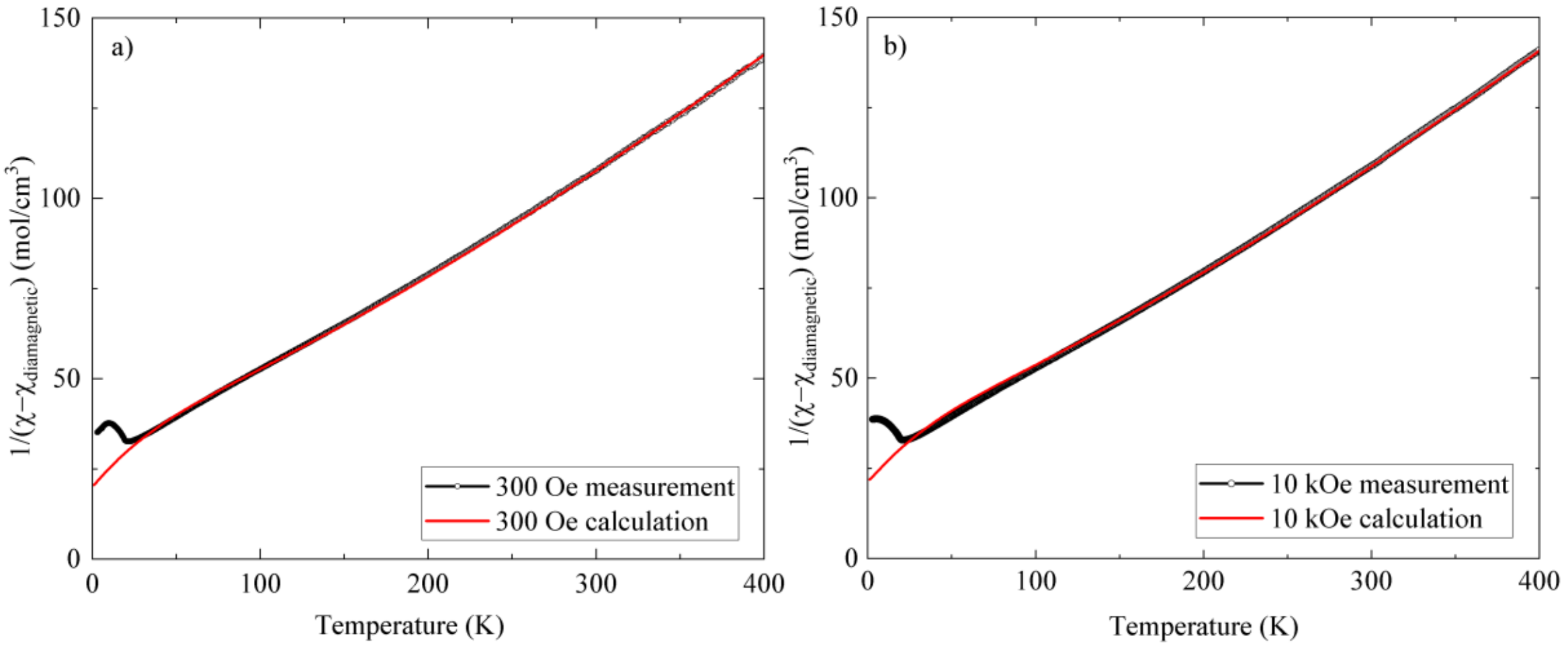

Figure 4. Temperature dependence of inverse magnetic susceptibility of BCMO samples at 0.3 kOe (a) and 10 kOe (b) compared with simulation based on cluster model derived with parameters derived from XAS measurements.

Isothermal magnetization measurements provide evidence for the slight anisotropy of BCMO, as seen from non-linear $M(H)$ responses and their field derivatives for crystals oriented along [100] and [111]. Figure 5 represents the magnetization curves and the first derivative of magnetization for the two crystalline samples at $T = 3$ K. The measurement revealed that the magnetization of BCMO has nonlinear behavior for the [100] and [111] crystals. The first derivative of magnetization of [100] and [111] shows a pronounced peak at 26.5 kOe consistent with a spin-flop transition. In a classical easy-axis two-sublattice antiferromagnet, a spin-flop transition occurs when the magnetic field ($H_{sf} = 26.5$ kOe) is applied along the easy axis; below $H_{sf}$, the sublattice moments are antiparallel and collinear with the field, whereas above $H_{sf}$ the staggered magnetization reorients to lie nearly perpendicular to the field, with a small canting toward $H$.
Neutron powder diffraction on BCMO has established a type-II antiferromagnetic structure with propagation vector $\mathbf{k} = (1/2,1/2,1/2)$, corresponding to ferromagnetic Co layers perpendicular to [111] stacked antiferromagnetically along this direction [22]. This suggests dominant antiferromagnetic second neighbor exchange interactions via 180° Co-O-Mo-O-Co pathways, while the frustrated first neigbor interaction is much weaker. Although the spin direction was not determined, this structure admits several symmetry-related domains with different equivalent orientations of the local easy-axis relative to the crystallographic axes. In our single crystal, both $H \parallel [100]$ and $H \parallel [111]$ have substantial projections onto the local easy axes of at least some domains, so a spin-flop/spin-reorientation transition appears in the magnetization for both field directions.

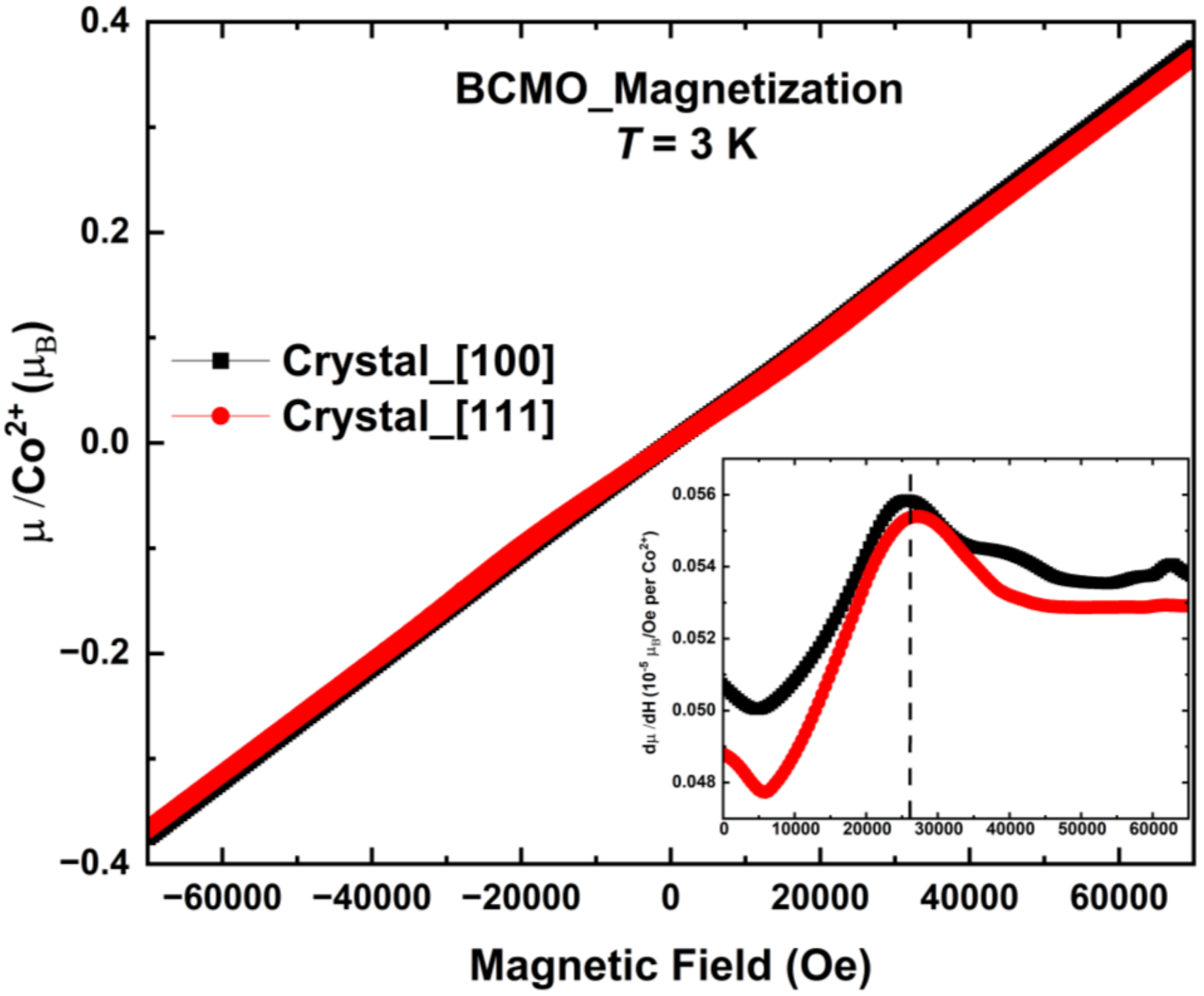

Figure 5. Isothermal magnetization of different BCMO crystals with field applied along the [100] and [111] directions at $T = 3$ K. The inset shows the derivative of the effective moment per $Co^{2+}$ ion for the directions [100] (black) and [111] (red). The dashed line represents a spin-flop transition around 26.5 kOe for both the [100] and [111] directions.

Very similar behavior is observed in isostructural $Ba_2CoWO_6$, where spin-flop transitions occur for $H \parallel [100]$ and $H \parallel [111]$, but not for $H \parallel [110]$, suggesting a common easy-axis orientation in the $Ba_2CoB''O_6$ (B″ = Mo, W) family [27]. The observation of a spin-flop transition implies a finite magnetic anisotropy, which stabilizes an antiferromagnetic ground state with a preferred spin orientation [28]. The spin-flop field reflects the balance between anisotropy energy and exchange interactions. Taken together, these results indicate that $Ba_2CoMoO_6$ is governed by antiferromagnetic exchange interactions and weak anisotropy.

## Heat Capacity

Figure 6 (a) shows the heat capacity of BCMO single crystals up to 50 K and of the polycrystalline boules up to 200 K, revealing a second-order anomaly at $T_N = 20$ K consistent with the onset of long-range antiferromagnetic order inferred from susceptibility data. No significant change was noticed in the heat capacity under different applied magnetic fields up to 90 kOe and the Néel temperature is almost field independent for all samples. The transition peak is noticeably sharper in the crystals than in powder, reflecting higher sample quality and reduced disorder broadening in the single crystals. To isolate the magnetic contribution, the lattice heat capacity of the polycrystalline sample was modelled between 50-170 K, well above the magnetic ordering temperature $T_N$, using a Debye–Einstein scheme with one Debye and one Einstein branch in a 51.5:48.5 weight ratio, according to the equation:

$$c_p\,(T) = 3Rn\left(0.515\left(\frac{T}{\theta_D}\right)^3 3\int_0^{\frac{\theta_D}{T}} \frac{x^4\, e^x}{(\,e^x - 1)^2}dx + 0.485\left(\frac{\theta_E}{T}\right)^2 \frac{e^{\theta_E/T}}{(e^{\theta_E/T} - 1)^2}\right),$$

where $R$ is the universal gas constant, $n$ is the number of ions per formula unit which is $n = 10$ for BCMO, and $\theta_D, \theta_E$ are the Debye and Einstein temperatures, respectively. The resulting fit yields $\theta_D = 277(4)$ K and $\theta_E = 568(5)$ K which provide a reasonable description of the phonon background over the chosen window and a stable baseline for extracting the magnetic heat capacity, $c_{\mathrm{mag}}$, by extrapolating to low temperature and subtracting from the measured data. The extracted magnetic heat capacity and entropy changes are shown in Figure 6(c). The weak drop in $c_{\mathrm{mag}}$/T between 60- 80 K is most likely due to the combined effects of phonon-subtraction uncertainty and residual impurities in the sample.

The entropy change $\Delta S$ was calculated by integrating $c_{\mathrm{mag}}/T$ in the range 2-100 K. The magnetic entropy saturates at ~50 K at the value 5.5(1) J mol$^{-1}$ K$^{-1}$or 0.95(2) $R\ ln\ 2$. The saturation value is consistent within uncertainty with a doublet ground state described by $J_{\mathrm{eff}}$ = ½, for which $\Delta S = R\ ln(2J + 1) = R\ ln\ 2$. This is similar to $Ba_2CoWO_6$ which also has $J_{\mathrm{eff}}$ = ½ [27]. 41% of the entropy is released above the Néel temperature up to 50 K. The extended release of entropy above $T_N$ suggests the persistence of correlated magnetic fluctuations well above $T_N$. The slightly reduced value of entropy change of 0.95(2) $R\ ln\ 2$ by $T = 50$ K could be attributed to a conservative phonon-background subtraction together with the minor secondary phases, resulting in slightly underestimation of the extracted entropy.

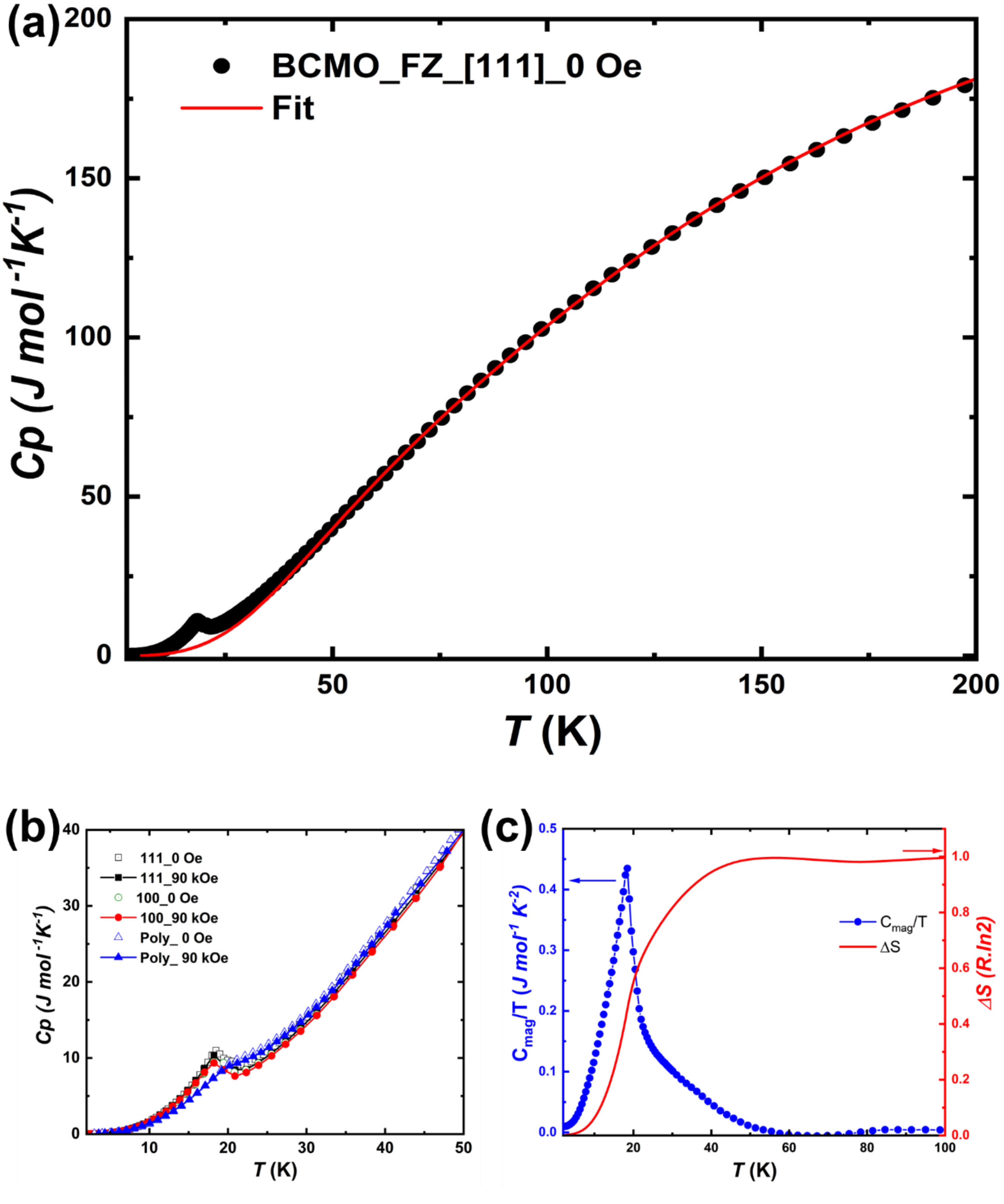

Figure 6. (a)The heat capacity of BCMO polycrystalline sample at zero field fitted by Debye and Einstein models to subtract the magnetic contribution of the heat capacity in the temperature range 50-150 K. (b) Heat capacity of BCMO crystals for magnetic field applied along the [100] and [111] directions and for the sol-gel polycrystals. (c) Magnetic contribution of the heat capacity over temperature (Left axis) for the polycrystalline sample in zero field. The entropy change is shown on the right axis in units of *R ln 2*.

## Electronic structure

### A. X-ray absorption spectroscopy

To probe the electronic and magnetic properties of BCMO we performed temperature dependent X-ray absorption spectroscopy (XAS) using the inverse partial fluorescence yield method [29,30], which is suitable for extremely insulating samples, in a similar manner to what is reported in Ref. [31]. XAS is extremely sensitive to the local electronic structure, and when combined with theoretical modelling to simulate the measured spectra, it is possible to obtain information about the ground state quantum numbers and effective crystal field parameters [14–17, 32, 33]. Part of the residual polycrystalline boules from CZ crystal growth was mounted in an aluminum sample holder and cleaved in ultra-high vacuum conditions before being transferred *in-vacuo* to the measurement chamber with a base pressure of ~$10^{-11}$ mbar.

Figure 7 shows the temperature-dependent isotropic Co $L_{2,3}$ XAS spectra of BCMO. In XAS the measured spectra are separated into the two $L_2$ and $L_3$ edges due to *2p* spin-orbit coupling. The obtained line shape is characteristic of a $Co^{2+}$ valence, which we will explain with our calculations. The temperature dependence originates from the thermal population of low-lying excited states with increasing temperature. This behavior is expected for high-spin $Co^{2+}$ ions in octahedral coordination [15].

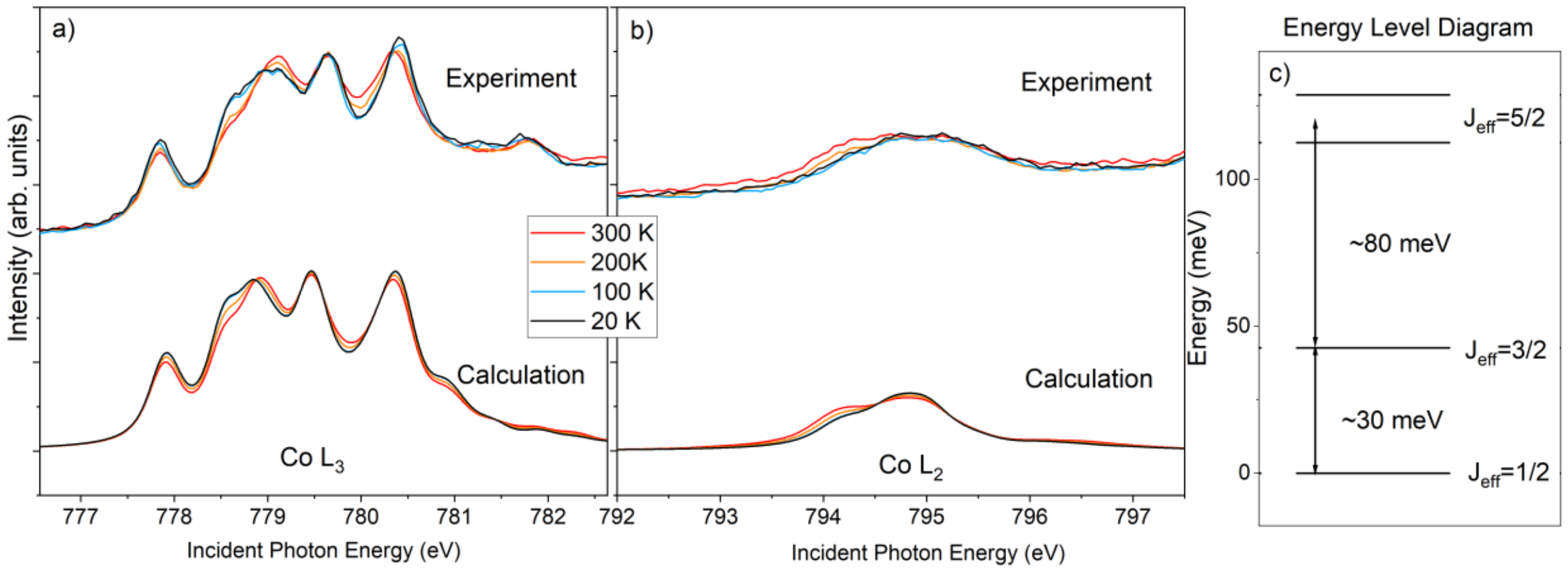

Figure 7. Experimental and calculated temperature dependence of the XAS spectra of $Ba_2CoMoO_6$ for the (a) $L_3$ and (b) $L_2$ Co edges and (c) the corresponding energy level diagram for the calculation without core-hole .

To obtain further information about the electronic structure of the compound, we performed full-multiplet configuration-interaction cluster calculations for a $CoO_6$ cluster using the $O_h$ point group. Hybridization parameters were obtained from Wannierized DFT band structure calculations [35] using the FPLO code [36]. The cluster-interaction calculations were performed with the Quanty code [35,37–39]. The full set of parameters used in our calculations is listed as Ref. [40]. Our calculations include *3d-3d* and *2p-3d* Coulomb interactions, spin-orbit coupling for both the *2p* and *3d* shells, hybridization with oxygen ligands and crystal-field effects. We obtain a very good agreement between experiment and calculation, implying that our model accurately captures most details of the local electronic structure of $Co^{2+}$ ions in this material. Importantly, the XAS results confirm that Co ions in this material are divalent, $Co^{2+}$. Charge neutrality then requires Mo to be in the hexavalent $Mo^{6+}$ oxidation state, consistent with the observed insulating character of the compound. Furthermore, we performed XPS analyses to detect any absorption from both $Mo^{5+}$ and $Mo^{6+}$ in the polycrystalline sample. The Mo 3d XPS spectrum shows a main doublet in the $Mo^{6+}$ binding-energy region. No clear separate feature attributable to $Mo^{5+}$ is observed in the raw spectrum, suggesting that Mo is predominantly in the hexavalent state (Additional details on the analysis of XPS measurements are provided in the Supplemental Material [53]).

The parameters that we have obtained from the XAS analysis (listed as Ref, 40) serve as a good starting point for describing the near ground state properties of the material. Now we need to consider that the presence of the core hole in the XAS process may affect the radial extent of the 3d orbital. This implies that the parameters in describing the electronic states in core-hole-free experiments may be somewhat different [41-43]. For example, the crystal field felt by the 3d electrons as well as the hybridization strength between the 3d and the ligands tend to be smaller in the presence of the core hole. To reproduce the magnetic susceptibility data, we have to increase the 10Dq value to 0.85 eV, the hybridization strength by 20%, and reduce the SOC parameter to 46 meV. With these adjustments we were able to obtain an excellent match with the experiment, see Fig, 4. We then estimate the *g*-factor using the relation $\Delta E = g\mu_B B$, where $\Delta E$ is the splitting of the doublet due to magnetic field. In a calculation with applied field of 6 T at 0 K, we obtain a value of $g = 4.36$.
From our model we find the first low lying spin-orbit coupling excited state at 43 meV, as shown in Figure 7 (c).

### B. Surface photovoltage spectroscopy

Surface photovoltage (SPV) spectroscopy is a contactless, surface-sensitive probe of illumination-induced changes in near-surface band bending and photocarrier separation, and thus reports on optically driven excitations in the vicinity of the surface [44,45]. In $Ba_2CoMoO_6$ (BCMO), the SPV spectrum (Fig. 8) exhibits a clear onset near 1.5 eV, a pronounced maximum at 2.65 eV, and a broad decay in intensity up to about 4.5 eV, showing that the near-surface photoresponse is dominated by a distinct transition manifold rather than a featureless background. Because SPV is inherently sensitive to surface and space-charge-region physics, the low-energy onset should be interpreted as an effective photoresponse threshold influenced by band bending and defect states, not as a direct measurement of the bulk band gap [44-46].
The strong SPV maximum at 2.65 eV is naturally associated with the dominant crystal-field/charge-transfer transition of the Co–O–Mo network on the face-centred-cubic lattice. Its energy scale is consistent with reported optical gaps in related Ba-based double perovskites: $Ba_2CoWO_6$ shows a cubic *Fm-3m* structure with an optical gap of about 2.4–2.6 eV from UV–visible and SPV analysis [27], while $Ba_2CaMoO_6$ and other $Ba_2AMoO_6$ compounds exhibit direct gaps in the range 2.3–2.8 eV from combined DFT and diffuse-reflectance measurements [47-49]. In this context, the 2.65 eV feature in BCMO fits naturally into the Mo/W double-perovskite landscape and indicates that the Co–O–Mo manifold hosts a robust, visible-range optical transition that efficiently drives surface photovoltage [50].
The SPV spectrum reflects multiple photoactive channels, arising from a combination of intrinsic electronic transitions and surface- or defect-related processes [44-

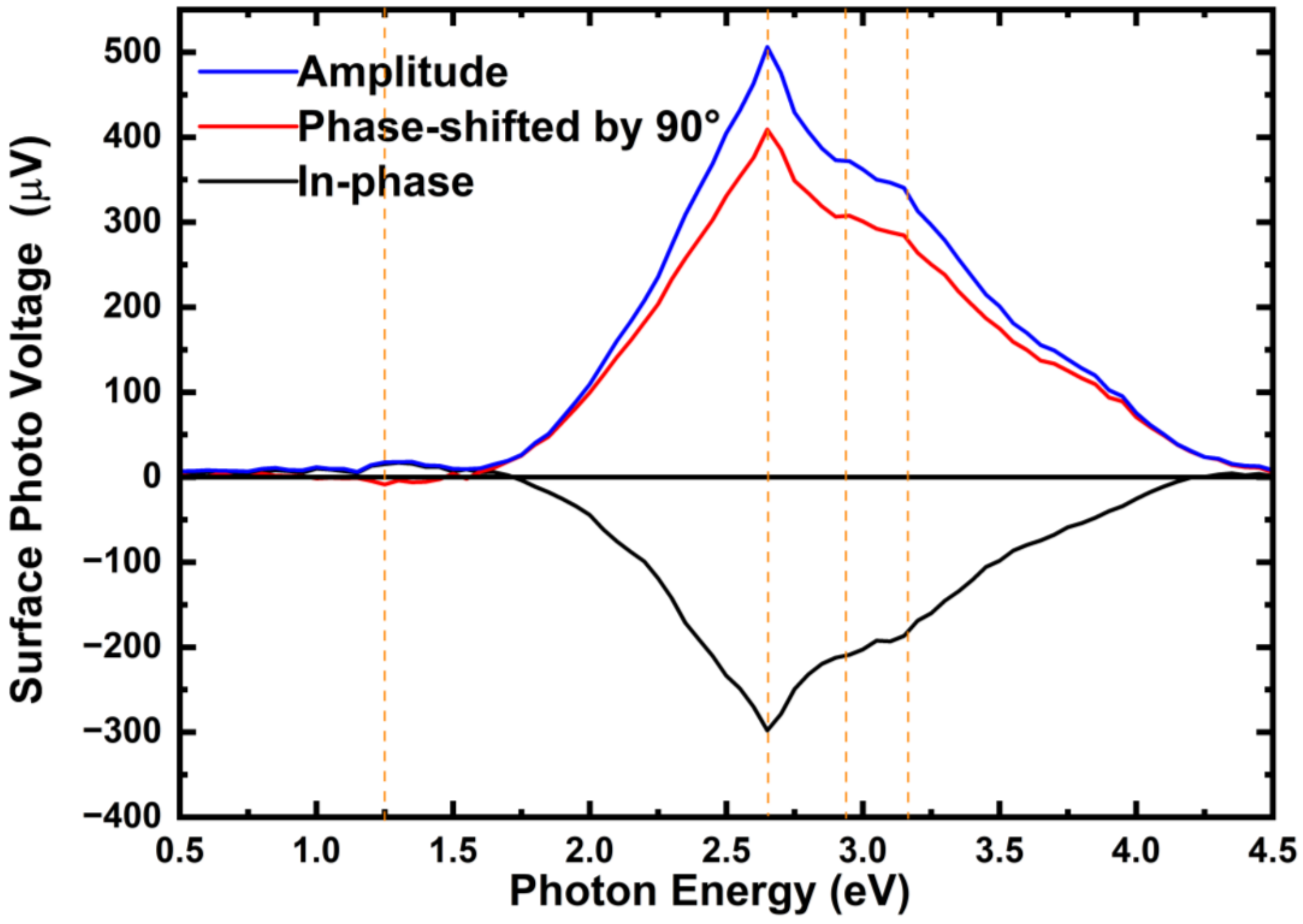

Figure 8. (a) Surface photovoltage (SPV) measurement of polycrystalline pellets of BCMO plotted as a function of photon energy: The in-phase signal (X component) (black), the phase-shifted by 90° signal (Y component) (red), and the SPV amplitude (composed of both X and Y components) (blue). The dashed orange lines represent the absorption of the BCMO phase.

45]. A more detailed separation of *d–d*, charge-transfer, and band-to-band contributions will require combining SPV with bulk-sensitive optical probes, such as ellipsometry or diffuse reflectance spectroscopy, and targeted first-principles calculations [49,50]. Within the present data, however, the robust SPV response and prominent 2.65 eV feature already demonstrate that BCMO hosts a strong, near-surface optoelectronic response in the visible–near-UV range, comparable in energy scale to $Ba_2CoWO_6$ and $Ba_2CaMoO_6$ but realized here in a spin–orbit-entangled FCC antiferromagnet. This combination of antiferromagnetic order, Co-based crystal-field physics, and efficient photovoltage generation highlights $Ba_2CoMoO_6$ as a promising platform for spintronic optoelectronic devices and photocatalytic or photoelectrochemical applications, where simultaneous control of magnetic order and interfacial charge separation is desired [50-52].

## Conclusions

This work establishes high-quality $Ba_2CoMoO_6$ (BCMO) single crystals as a robust platform for studying intrinsic magnetic and electronic properties of face-centered cubic (FCC) antiferromagnets with strong spin–orbit coupling. Optimized floating-zone and top-seeded solution growth under argon atmosphere produced crystals with reduced $BaMoO_4$ impurity fractions compared to the powder samples, confirming the cubic *Fm-3m* structure with lattice parameter a = 8.088 Å. Rietveld refinements of powder x ray diffraction data, including the superlattice reflections associated with B site order, show that Ba, Co, and Mo occupancies remain close to ideal stoichiometry and that oxygen occupancy is stoichiometric within experimental uncertainty, indicating rock salt ordering of Co and Mo and no measurable antisite disorder or mixed valences as supported by XPS spectroscopy.
Magnetic susceptibility and heat capacity measurements consistently reveal long-range antiferromagnetic ordering at $T_N$ = 20.1(1) K, which is unaffected by applied fields up to 90 kOe. Quantitative analysis of the [111] single-crystal susceptibility using the cluster model derived from XAS analysis reproduces the measured curve. The fitted molecular-field parameters gives an effective Weiss Temperature -37 K. Isothermal magnetization reveals a spin-flop transition at 26.5 kOe for both the [100] and [111] orientations, indicating weak magneto-crystalline anisotropy and ordered-moment components along these crystallographic directions.
The Debye–Einstein fitting of the heat capacity provides the phonon background needed to isolate the magnetic contribution $c_{\mathrm{mag}}$, from which the magnetic entropy is obtained by integrating $c_{\mathrm{mag}}/T$. The fact that the entropy saturates close to *R ln 2* indicates that, over the measured temperature range, the thermally accessible magnetic degrees of freedom correspond to a two-level manifold. For $Co^{2+}$ in the present octahedral environment, this is consistent with a Kramers doublet ground state produced by the combined action spin-orbit coupling and crystal-field splitting, in agreement with the $J_{\mathrm{eff}}$=½ state inferred from XAS. X-ray absorption spectroscopy independently verifies the $Co^{2+}$ valence state, the local octahedral crystal-field scheme, the doubly degenerate ground state with $J_{\mathrm{eff}}$ = ½ character and yields a *g*-factor *g* = 4.36, consistent with the enhanced moment observed in susceptibility. For Mo, XPS Mo-edge spectroscopy indicates $Mo^{6+}$ as supported by charge balance, the $Co^{2+}$ assignment, and the absence of reduced Mo phases in the diffraction data. Surface photovoltage spectroscopy suggests a strong optical response linked to crystal-field and band-structure transitions, implying potential for optoelectronic applications, though bulk optical measurements (ellipsometry, diffuse reflectance) would complement this surface-sensitive SPV data.

To summarize, these results position $Ba_2CoMoO_6$ alongside $Ba_2CoWO_6$ as a model weakly frustrated FCC antiferromagnet, distinguished by its robust $J_{\mathrm{eff}}$ = ½ ground state, modest anisotropy, and accessible spin-flop behavior. The weak-frustration assignment is suggested by the release of 41% of the expected entropy above the Neel temperature, the effective Weiss temperature of -37 K , as well as the microscopic description of the magnetic response provided by the XAS-derived cluster mode. Future work will combine single-crystal neutron diffraction, bulk optical spectroscopy, and first-principles calculations. This approach will refine the magnetic structure and clarify electronic-level assignments. Additionally, assess spintronic and energy-conversion applications which are suggested by the spin-flop transition and strong photovoltage response. The transition enables field-controlled reorientation of the Néel vector. Consequently, this mechanism may strongly modulate transport signals.

## Conflict of interest

The authors declare no conflict of interest.

## Acknowledgements

We acknowledge the CoreLab Quantum Materials at Helmholtz Zentrum Berlin für Materialien und Energie (HZB), Germany, where the powder and single crystal samples of $Ba_2CoMoO_6$ were synthesized and characterized. The authors acknowledge the support of Igal Levine and Thomas Dittrich from the Solar Energy Division, HZB, for performing the surface photovoltage measurements. We acknowledge Christian Höhn from the Institute of Solar Fuels, HZB for performing the X-ray photoemission spectroscopy. M.M.F.-C. greatly acknowledges funding from the Deutsche Forschungsgemeinschaft (DFG, German Research

Foundation) Grant No. 387555779. Work at MPI-CPfS Dresden and HZB was partially supported by SFB1143 (Project No. 247310070). The authors acknowledge the support from the Max Planck-POSTECH-Hsinchu Center for Complex Phase Materials.

**Supporting Information**

Supporting Information available for this article includes additional magnetic characterization of $Ba_2CoMoO_6$, comprising a Lloret-model treatment of the [111] crystal to account for the spin–orbit-coupled $Co^{2+}$ response. It also includes XPS measurements and core-level fitting of the Mo 3d region, which establish the exclusively $Mo^{6+}$ valence state. The relevant data are presented in Figs. SI-1–SI-3 and Tables SI-1–SI-2.

**Single-Crystal Growth and Physical Properties of the Antiferromagnetic Double Perovskite $Ba_2CoMoO_6$**

A.R.N. Hanna[1,*], M. M. Ferreira-Carvalho[2,3], S.H. Chen[2], C. F. Chang[2], C. Y. Kuo[4,5], A.T.M.N. Islam[1], R. Feyerherm[1], L.H. Tjeng[2] , B. Lake[1,6,*]

[1]Helmholtz-Zentrum Berlin für Materialien und Energie GmbH, 14109 Berlin, Germany

[2]Max Planck Institute for Chemical Physics of Solids, Nöthnitzer Str. 40, 01187 Dresden, Germany

[3]Institute of Physics II, University of Cologne, Zülpicher Straße 77, 50937 Cologne, Germany

[4]Department of Electrophysics, National Yang Ming Chiao Tung University, 30010 Hsinchu, Taiwan

[5]National Synchrotron Radiation Research Center, 101 Hsin-Ann Road, 30076 Hsinchu, Taiwan

[6]Institut für Festkörperphysik, Technische Universität Berlin, Hardenbergstraße 36, D-10623 Berlin, Germany

*Corresponding authors : Abanoub.Hanna@helmholtz-berlin.de, Bella.lake@helmholtz-berlin.de

## Magnetic Properties

## Lloret model

To describe the $Co^{2+}$ response quantitatively within an analytical framework, the susceptibility of the ⟨111⟩ crystal at 300 Oe was also analyzed using the model developed for six-coordinate high-spin $Co^{2+}$ complexes by *Lloret et al.* [26], as shown in Fig. SI-1. In this approach, the temperature and spin-orbit dependence of the effective moment $\mu_{eff}$(T,λ) is encoded in closed-form expressions obtained for the $^4T_1$(F) ground term in an octahedral crystal field. The susceptibility data were expressed using the equation:

$$\chi_{mol} = \chi_0 + \frac{N_A\,\mu_0 {\mu_B}^2}{3k_B(T - \theta_{CW}\,)}\mu_{eff}^2(T,\lambda)$$

with

$$\mu_{eff}^2(T,\lambda) = 3\,\frac{c_1 + c_2\frac{k_BT}{\lambda} + (c_3 + c_4\frac{k_BT}{\lambda})\,exp\left(\frac{9\lambda}{4k_BT}\right) + (c_5 + c_6\frac{k_BT}{\lambda})\,exp\left(\frac{6\lambda}{k_BT}\right)}{1 + 2\,exp\left(\frac{9\lambda}{4k_BT}\right) + 3\,exp\left(\frac{6\lambda}{k_BT}\right)}.$$

Here, $\chi_0$ is the temperature-independent contribution, $\theta_W$ is the effective Weiss temperature, $\lambda$ is the spin–orbit coupling parameter, and the coefficients $c_1$–$c_6$ are fixed

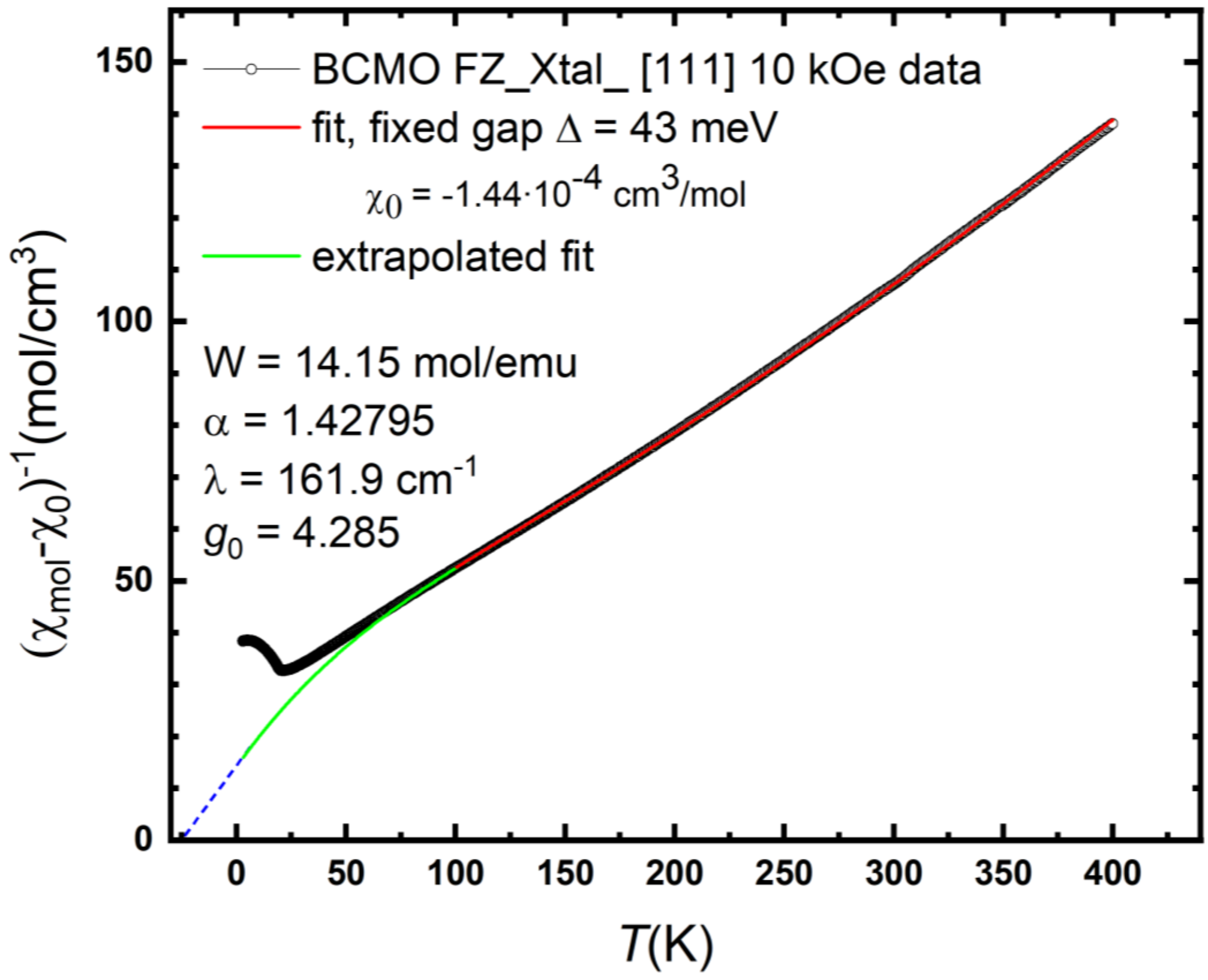

Figure SI-1. Temperature dependence of the molar magnetic susceptibility, $\chi_{\mathrm{mol}}$, and inverse susceptibility, $(\chi_{\mathrm{mol}} - \chi_0)^{-1}$, of the BCMO FZ crystal measured at 0.3 kOe with the magnetic field applied along the $\langle 111 \rangle$direction. The red line represents a fit using the Lloret model for six-coordinate high-spin $Co^{2+}$, with the spin–orbit gap fixed to $\Delta = 43$ meV, corresponding to $|\lambda| = 154.1\ \mathrm{cm}^{-1}$. The green line shows the extrapolated fit.

numerical constants tabulated by *Lloret* et al. [26]. The effective Weiss temperature was calculated from

$$\theta_{\mathrm{W}} = \frac{g^2}{4}\frac{N_{\mathrm{A}}\mu_{\mathrm{B}}^2}{k_{\mathrm{B}}}W,$$

with

$$g = \frac{10 + 2\alpha}{3}, \frac{N_{\mathrm{A}}\mu_{\mathrm{B}}^2}{k_{\mathrm{B}}} = 0.37515\ \mathrm{emu\ K\ mol^{-1}}.$$

The spin–orbit gap was fixed to $\Delta = 43$ meV, as determined independently from the XAS analysis. Using $\Delta = \frac{9|\lambda|}{4}$, which corresponds to $|\lambda| = 154.1\ \mathrm{cm}^{-1}$. The temperature-independent diamagnetic susceptibility was calculated from the diamagnetic correction of each element as $\chi_0 = -1.44 \times 10^{-4}\ \mathrm{cm}^3\ \mathrm{mol}^{-1}$ and was included as a constant in the model [54].

Therefore, $\lambda$ was not varied independently, and the fit contained only two free parameters: $\alpha$ and $W$. The fitted values were $\alpha = 1.42795, g = 4.285$ *and* $W = 14.15$ mol/emu. These parameters yield $\theta_W = -24.35$ K.

The effective Weiss temperature $\theta_W$ obtained from the Lloret molecular-field parameter $W$ accounts for the interionic magnetic interactions within the Weiss molecular-field approximation, while $\chi_0$ accounts for temperature-independent contributions. The Lloret model is used as a phenomenological description of the single-ion $Co^{2+}$ response; the microscopic interpretation of the magnetic properties is based on the XAS-derived cluster model discussed in the main text.

## X-ray photoelectron spectroscopy (XPS)

X-ray photoelectron spectroscopy (XPS) analyses were performed at room temperature using a monochromatic Al Kα X-ray source (Focus 500, Specs) and a Phoibos 100 electron analyzer (Specs) at the Institute for Solar Fuels, Helmholtz-Zentrum Berlin (HZB), as shown in Fig. SI-2(a). XPS was utilized to investigate the presence of different molybdenum oxidation states (namely $Mo^{5+}$ and $Mo^{6+}$ ions) within the polycrystalline BCMO boules. High-resolution spectra of the Mo 3d core level were deconvoluted and peak-fitted using CasaXPS software [55]. Due to sample charging induced by the insulating nature of the specimen mounted on conductive carbon tape, the adventitious C 1s peak—representing the aliphatic C–C/C–H backbone overlaid with the oxygenated C–O components of the tape adhesive matrix—was initially recorded at 286.00 eV. To correct for this surface charging effect, a global binding energy shift of -1.00 eV was applied across all core levels, calibrating the reference aliphatic C 1s position to the standard value of 285.00 eV [56].
Following this charge calibration, the raw experimental Mo 3d envelope was successfully fitted using a single spin-orbit doublet assigned exclusively to the $Mo^{6+}$ chemical state, as shown in Fig. SI-2(b). The high-resolution Mo 3d spectrum was deconvoluted with symmetric GL(30) component line shapes utilizing a physically derived Tougaard background to accurately model the inelastic electron scattering profile of the insulating $Mo^{6+}$ matrix. The core Mo $3d_{5/2}$ and Mo $3d_{3/2}$ peaks were positioned at 232.47 eV and 235.74 eV, respectively, matching established literature values for fully oxidized molybdenum systems [57]. The complete deconvolution metrics, including binding energies, peak areas, and relative concentrations, are summarized in Table SI-1. To ensure a physically meaningful fit, synthetic component constraints were applied: the Full Width at Half Maximum (FWHM) values of the two components were constrained to be equal to 1.82 eV, and the spin-orbit splitting energy separation (ΔE) was maintained at 3.07 eV [57, 58]. Furthermore, the relative area ratio of the $3d_{5/2}$ to $3d_{3/2}$ components yielded a 3:2 ratio (60.06% to 39.94%), strictly conforming to the theoretical *2j+1* multiplicity requirement for d-orbitals. These results indicate that Mo is predominantly present as $Mo^{6+}$ in the near-surface region probed by XPS, with no detectable $Mo^{5+}$ or $Mo^{4+}$ components within the experimental detection limit.

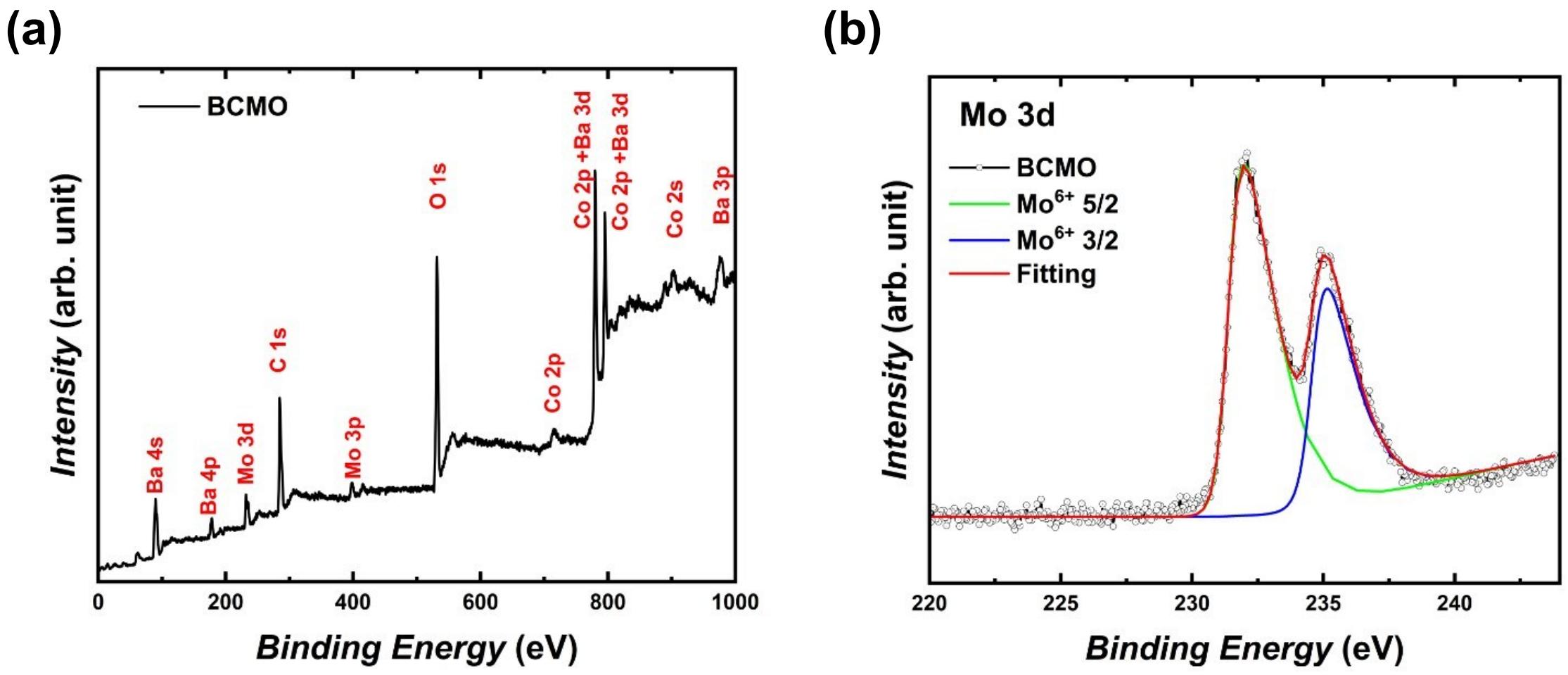

*Figure SI-2: The XPS spectrum of BCMO polycrystalline boules. (a) The XPS survey spectrum indicates the main photoelectron peaks attributed to the elements of BCMO and carbon substrate. All data are charge-corrected to the reference adventitious C 1s position at 285.00 eV(b)High-resolution X-ray photoelectron spectrum and corresponding peak deconvolution of the Mo 3d core level for polycrystalline BCMO. The experimental data (scatter points, black) are fitted with a single spin-orbit doublet (solid lines, green, red, blue) representing the $Mo^{6+}$ state (Mo $3d_{5/2}$ at 232.47 eV and Mo $3d_{3/2}$ at 235.54 eV), confirming the absence of lower $Mo^{5+}$ or $Mo^{4+}$ oxidation states.*

*Table SI-1. Detailed core-level peak fitting parameters and relative atomic concentrations for the Mo 3d region of the polycrystalline BCMO boule.*

| *Component* | Orbital Region | Binding Energy (eV) | FWHM (eV) | Raw Area (cps·eV) | Relative Atomic Conc. (%) |
|---|---|---|---|---|---|
| *Total Mo 3d* | Envelope | 232.28 | 2.08 | 3,742.07 | 100.00 |
| *$Mo^{6+}$ $3d_{5/2}$* | Component | 232.47 | 1.82 | 2,135.43 | 60.06 |
| *$Mo^{6+}$ $3d_{3/2}$* | Component | 235.54 | 1.82 | 1,423.62 | 39.94 |